\pdfoutput=1
\documentclass[numberedappendix]{emulateapj-rtx4}
\usepackage{natbib}
\usepackage{psfig}
\usepackage{amsmath}

\newcommand{\pdif}[2]{\ensuremath{ \frac{\partial #1}{\partial #2}}}

\shorttitle{Simulations of Dust-Driven winds}
\shortauthors{Zhang \& Davis}

\begin{document}

%% LaTeX will automatically break titles if they run longer than
%% one line. However, you may use \\ to force a line break if
%% you desire.

\title{Radiation Hydrodynamic Simulations of Dust-Driven Winds}

%% Use \author, \affil, and the \and command to format
%% author and affiliation information.
%% Note that \email has replaced the old \authoremail command
%% from AASTeX v4.0. You can use \email to mark an email address
%% anywhere in the paper, not just in the front matter.
%% As in the title, use \\ to force line breaks.

\author{Dong Zhang\altaffilmark{1} and Shane W. Davis\altaffilmark{1}}
\affil{Department of Astronomy, University of Virginia, Charlottesville, VA 22904, USA}
\email{dz7g@virginia.edu}

%%\author{C. D. Biemesderfer\altaffilmark{4,5}}
%%\affil{National Optical Astronomy Observatories, Tucson, AZ 85719}
%%\email{aastex-help@aas.org}

%%\and

%%\author{R. J. Hanisch\altaffilmark{5}}
%%\affil{Space Telescope Science Institute, Baltimore, MD 21218}

%% Notice that each of these authors has alternate affiliations, which
%% are identified by the \altaffilmark after each name.  Specify alternate
%% affiliation information with \altaffiltext, with one command per each
%% affiliation.

%%\altaffiltext{1}{Visiting Astronomer, Cerro Tololo Inter-American Observatory.
%%CTIO is operated by AURA, Inc.\ under contract to the National Science Foundation.}
%%\altaffiltext{2}{Society of Fellows, Harvard University.}
%%\altaffiltext{3}{present address: Center for Astrophysics, 60 Garden Street, Cambridge, MA 02138}
%%\altaffiltext{4}{Visiting Programmer, Space Telescope Science Institute}
%%\altaffiltext{5}{Patron, Alonso's Bar and Grill}

%% Mark off your abstract in the ``abstract'' environment. In the manuscript
%% style, abstract will output a Received/Accepted line after the
%% title and affiliation information. No date will appear since the author
%% does not have this information. The dates will be filled in by the
%% editorial office after submission.

\begin{abstract}

We study dusty winds driven by radiation pressure in the atmosphere of a rapidly star-forming environment. We apply the variable Eddington tensor algorithm to re-examine the two-dimensional radiation hydrodynamic problem of a column of gas that is accelerated by a constant infrared radiation flux. In the absence of gravity, the system is primarily characterized by the initial optical depth of the gas. We perform several runs with different initial optical depth and resolution. We find that the gas spreads out along the vertical direction, as its mean velocity and velocity dispersion increase. In contrast to previous work using flux-limited diffusion algorithm, we find little evolution in the trapping factor. The momentum coupling between radiation and gas in the absence of gravity is similar to that with gravity. For Eddington ratio increasing with the height in the system, the momentum transfer from the radiation to the gas is not merely $\sim L/c$, but amplified by a factor of $1+\eta \tau_{\rm IR}$, where $\tau_{\rm IR}$ is the integrated infrared optical depth through the system, and $\eta\sim0.5-0.9$, decreasing with the optical depth. We apply our results to the atmosphere of galaxies and conclude that radiation pressure may be an important mechanism for driving winds in the most rapidly star-forming galaxies and starbursts.

\end{abstract}

%% Keywords should appear after the \end{abstract} command. The uncommented
%% example has been keyed in ApJ style. See the instructions to authors
%% for the journal to which you are submitting your paper to determine
%% what keyword punctuation is appropriate.


\section{Introduction}

Feedback processes play a crucial role in galaxy formation and evolution. In particular, radiation pressure from the continuum absorption and scattering of starlight on dust grains has been proposed as an important mechanism in driving supersonic turbulence in the interstellar medium (ISM), hampering gravitational collapse, and launching large-scale galactic winds in starbursts and rapidly star-forming galaxies. One-dimensional analytic models show that dusty winds can be driven by radiation pressure in rapidly star-forming environments, such as luminous infrared galaxies (LIRGs) and ultraluminous infrared galaxies (ULIRGs) (e.g., \citealt{Thompson05}; \citealt{Murray05}; \citealt{Murray11}; \citealt{Zhang12}).

However, the simplified galactic wind models contain some uncertainties. A key question that cannot be addressed by analytic models is how much momentum is coupled between radiation and dusty gas. In the single scattering limit, i.e., the system is optically thick to the UV photons, but optically thin to the re-radiated infrared (IR) emission from dust grains, all photons are absorbed and scattered once, the radiation transfers a momentum flux of $L/c$ in the gas, where $L$ is the luminosity of radiation source. However, it is uncertain how much momentum is transferred from radiation to gas if the system is optically thick to its own infrared emission. It has been argued that the rate of momentum deposition will never exceed a few of $L/c$ (\citealt{Krumholz09}), or it approaches $\tau_{\rm IR} L/c$, where $\tau_{\rm IR}\gg1$ is the mean IR optical depth of the system (\citealt{Thompson05}; \citealt{Murray10}; \citealt{Andrews11}; \citealt{Thompson15}). In order to understand the dynamics of radiation-gas interaction, multidimensional radiation hydrodynamics simulations have been carried out recently. 

%The momentum coupling between strong radiation forces in the radiation-dominated regime and the ISM has been studied recently by multidimensional radiation hydrodynamics simulations. 

Krumholz \& Thompson (2012, hereafter KT12) used a 2-dimensional (2D) model to investigate the efficiency of momentum transfer from IR opticlaly thick ultraluminous infrared galaxies (ULIRGs) to a dusty atmosphere with a vertically stratified gravity. Using 2D grey flux-limited diffusion (FLD) approximation in the \textsc{orion} code \citep{Krumholz07}, KT12 showed that the radiation-gas interaction gives rise to radiative Rayleigh-Taylor instability (RTI), driving supersonic turbulence, and limiting momentum transfer from the radiation to the gas to $\sim L/c$. The radiation momentum deposition in the regime where is initially sub-Eddington for dust is not sufficient to driven an unbound wind, most of the gas is eventually settled in a turbulent steady state confined near the base of the system. \citet{Skinner2015} reached a similar conclusion in a study of radiative feedback from a protocluster on a surrounding molecular cloud using their M1 closure method.

Using their variable Eddington tensor (VET) algorithm implemented in the \textsc{athena} code (\citealt{Stone08}; \citealt{Davis12}; \citealt{Jiang12}), Davis~ et al.(2014, hereafter D14) revisited the results of KT12 with the same 2D (and extented them to 3D). The VET algorithm is used to calculate the local Eddington tensor by solving the radiative transfer equation with the method of short characteristics (\citealt{Davis12}). In contrast to KT12, D14 showed a stronger momentum coupling between radiation and dusty gas. Although the radiative RTI develops and limit the radiation-gas interaction, The gas can be heated and accelerated upward by radiation, and produce an unbound outflow even from an initially sub-Eddington system. D14 showed that the significant difference between the outcome of simulations in KT12 and D14 resulted from limitations in the diffusion-based FLD scheme. The FLD and VET schemes agree well in the dense gas with optical depth $\tau_{\rm IR}\gg 1$, but the FLD approximation becomes inaccurate in modeling the radiation field responds to structure in the gas distribution in the system of $\tau_{\rm IR}\lesssim$ few.

\cite{Rosdahl15} simulated the same problem of KT12 and D14 using their new \textsc{ramses-rt} code with the M1 closure for the Eddington tensor. The M1 results show that the gas receives a larger acceleration than in the FLD calculations and reaches a large height, but this is ultimately insufficient to overcome the gravity and the gas eventually settles down in a marginally bound system, similar to the FLD results.  Hence, their results are qualitatively closer to those obtained with the FLD rather than with the VET method.  On the other hand, more recent simulations based on the implicit Monte Carlo radiation transfer scheme is more consistent with D14 (\citealt{Tsang15}). Both the M1 closure and FLD schemes impose artificial constraints on the radiation flow in optically thin regions while Monte Carlo and VET directly model the angular distribution of the radiation field.  The agreement between VET and the Monte Carlo algorithms along with the D14 analysis of how the FLD algorithm breaks down in optically thin regimes suggest that these algorithms are giving the most accurate representation of the flow for this problem setup.

Note that in the previous mentioned simulations of radiation-gas interaction, the size of the computational box is only about $\sim0.3$ pc$\times 1.3$ pc, with a resolution of $\Delta x \simeq 3.2\times 10^{-4}\,$pc, so that one can resolve the sound crossing timescale and the scale of gas turbulence. In order to investigate the efficiency of momentum coupling and wind propagation on a larger scale, Krumholz \& Thompson (2013, hereafter KT13) assumed that a wind is initially launched at the base of the galactic atmosphere due to super-Eddington radiation forces or other mechanisms, and turned off the gravity to study the maximum velocity the gas can gain from radiation. Using also \textsc{orion} and the FLD scheme, KT13 found that after wind acceleration begins, RTI forces the gas into a configuration that reduces the rate of momentum transfer from the radiation filed to the gas by a factor of $\sim10-100$ compared to an estimate based on the optical depth at the base of the atmosphere, the momentum transfer to gas is only a few of $\sim L/c$, without significant amplification by radiation trapping. They concluded that radiation pressure on dust is unlikely to be able to drive winds and ejecta from star-forming clusters and galaxies. So far, no other simulations have been done for the wind-gas interaction problem. Given previous discrepancies, it is important to re-examine the results of KT13 using the VET method.
 
%A further important question is, what is the momentum coupling between radiation and dusty gas, if a wind is really launched. Krumholz \& Thompson (2013, hereafter KT13) assumed the wind is launched by super-Eddington radiation forces or other mechanisms, and investigate the efficiency of momentum transfer using FLD method. They found that after wind accleration begins, radiation Rayleigh-Taylor instability forces the gas into a configuration that reduces the rate of momentum transfer from the radiation field to the gas by a factor of $\sim10-100$ compared to an estimate based on the optical depth at the base of the atmosphere. It is important to revisit the results of KT13 using VET method. 

This paper is organized as follows. In Section \ref{setup} we briefly summarize the equations and the simulation setup. The initial conditions of the gas are given by the end states of the simulations from D14. In Section \ref{results} we show our simulation with various parameters, and summarize our results. The astrophysical implications are discussed in Section \ref{sec_discussion}. Conclusions are given in Section \ref{conclusions}.

\section{Equations and Simulation Setup}\label{setup}

\subsection{Equations}

As in D14, we solve the equations of radiation hydrodynamics using \textsc{athena} with the built-in radiation module (\citealt{Davis12}; \citealt{Jiang12}). The equations of mass, momentum, energy, radiation energy and radiation momentum conservation are
\begin{eqnarray}
&&\pdif{\rho}{t} + \mathbf{\nabla} \cdot \left(\rho \mathbf{v} \right) = 0, \\
&&\pdif{\left(\rho\mathbf{v}\right)}{t} + \mathbf{\nabla} \cdot \left( \rho\mathbf{v} 
\mathbf{v} + {\sf P}\right) = \rho \mathbf{g} - \mathbf{S}_r(\mathbf{P}), \\
&&\pdif{E}{t} + \mathbf{\nabla} \cdot \left(E \mathbf{v} + {\sf P} \cdot \mathbf{v}\right) = 
\rho \mathbf{g} \cdot \mathbf{v} -c S_r(E), \\
&&\pdif{E_r}{t} + \mathbf{\nabla} \cdot \mathbf{F}_r=cS_r(E), \\
&&\frac{1}{c^2}\pdif{\mathbf{F}_r}{t}+\mathbf{\nabla} \cdot{\sf P}_r=\mathbf{S}_r(\mathbf{P}),
\end{eqnarray}
where $\rho$, $\mathbf{v}$, $\mathbf{g}$ are the gas density, fluid velocity and the gravitational acceleration, ${\sf P}=p{\sf I}$ is the pressure tensor, $p=\rho k_{B}T_g/\mu m_{\rm H}$ is the gas pressure, ${\sf I}$ is the identity matrix, and $E=p/(\gamma-1)+\rho v^{2}/2$ is the total fluid energy density. 

The radiation momentum and energy source terms $\mathbf{S}_r(\mathbf{P})$ and $S_r(E)$ are given by (\citealt{Lowrie99})
\begin{eqnarray}
\mathbf{S}_r(\mathbf{P}) & = &-\frac{\sigma_{F}}{c}\left[\mathbf{F}_r-
\left(\mathbf{v} E_r+\mathbf{v} \cdot {\sf P}_r\right)\right]  \nonumber \\ 
& &+\frac{\mathbf{v}}{c}(\sigma_{\rm P}a_rT^4-\sigma_{E}E_r),\label{source1}
\end{eqnarray}
\begin{eqnarray}
S_r(E) & = & (\sigma_{\rm P}a_rT^4-\sigma_{E}E_r)  \nonumber \\
& &+\sigma_{F}\frac{\mathbf{v}}{c^2}\cdot\left[\mathbf{F}_r-
\left(\mathbf{v} E_r+\mathbf{v}\cdot{\sf P} _r\right)\right],\label{source2}
\end{eqnarray}
where $E_r$ and $\mathbf{F}_r$ are the radiation energy density and radiation flux, $T$ is the gas temperature, $\sigma_F$, $\sigma_{\rm P}$ and $\sigma_{E}$ are the flux mean opacity, the Planck mean opacity and the energy mean opacity correspondingly, and $a_r$ is the Stefan-Boltzmann constant. For simplicity, we assume that the gas and the dust share a common temperature as $T$, and use the Planck $\kappa_{\rm P}$ and Rosseland $\kappa_{\rm R}$ mean opacities (KT12, KT13, D14)
\begin{eqnarray}
(\kappa_{\rm P}, \kappa_{\rm R}) = (10^{-1},10^{-3/2}) \left( \frac{T}{10 \; \textrm{K}}\right)^2 \; {\rm cm^2 \; g^{-1}}\label{opacity}.
\end{eqnarray}
Equation (\ref{opacity}) gives an approximation with a dusty gas at $T\lesssim 150\,$K (\citealt{Semenov03}). We set $\sigma_F=\rho \kappa_R$ in equations (\ref{source1}) and (\ref{source2}).

The thermal and dynamical behaviors of dust and gas has been discussed in KT13 (see their Appendix A). In the parameter space we are concerned with, the rate of dust-radiation energy exchange is higher than the rate of dust-gas energy exchange, therefore we expect that the dust is in thermal equilibrium with the radiation field, and we have the dust temperature $T_{\rm dust}\simeq T_r = (E_{r}/a_{r})^{1/4}$, where $T_r$ is the characteristic radiation temperature. On the other hand, the gas may have different temperature. \cite{Goldsmith01} showed that the dust thermally couples with the gas and has the same temperature only if the gas density exceeds $\sim10^{4}-10^{5}$ cm$^{-3}$. As the ISM material is accelerated and spreads out, the density of the gas drops quickly, and the gas no longer holds the same temperature as the dust, although they are still dynamically well coupled (\citealt{Murray05}; KT13). However, even in the case, the assumption $T_{\rm gas}\simeq T_{\rm dust}$ still provides a reasonable approximation. Since the gas is highly supersonic in the outflow, the thermal pressure of the gas is much weaker compared to ram pressure, changing the gas temperature is unlikely to significantly affect the dynamics of the gas. For simplification we still assume $T_{\rm gas}=T_{\rm dust}=T$ in our work. As discussed in D14, alternative simulations with $\kappa_{\rm R,P} \propto T_r$ were rung for the problem setup using non-zero gravity and the results agreed very closely with the simulations with $\kappa_{\rm R,P}\propto T$. Given that setting gravity to zero is the only significant change in the current setup, we assume this will still hold.  As the dusty neutral gas is accelerated farther from its origin, it likely to become more diffuse and some of the neutral gas will become more highly ionized, and the dust will be sublimated. Hence, our results here apply to an earlier neutral and dynamically well coupled phase of the outflow.

The radiation pressure ${\sf P}_r$ is given by ${\sf P} _r = {\sf f}E_r$, where ${\sf f}$ is the eponymous VET, which is calculate directly by
\begin{equation}
{\sf f}=\frac{{\sf P}_r}{E_r}=\frac{\int I(\hat{n}) \mu_i \mu_j d\Omega}{\int I(\hat{n}) d\Omega},
\end{equation}
where $d\Omega$ is the differential of solid angle, and $\mu_i \equiv \hat{n} \cdot \hat{x}_i$ is the cosine factor. The specific intensity of the radiation field $I$ can be solved by the radiation transfer equation
\begin{equation}
\hat{n} \cdot \nabla I = \sigma_{F} \left(\frac{a_r c}{4\pi} T^4 - I\right).
\label{eq:radtrans}
\end{equation} 
This equation is solved using the short characteristics method, as descried in detail in \cite{Davis12}.

\subsection{Dimensionless Units}

We define a constant flux $F_*$ as the source of the radiation field in the system injecting at the lower boundary, and 
\begin{equation}
T_*=\left(\frac{F_*}{a_r c}\right)^{1/4}
\end{equation}
is the characteristic temperature. Here, we follow the KT12 convention of denoting fiducial quantities with a ``*''.  Following KT12 and D14, we choose $T_*=82$\,K, which corresponds to $F_*=2.54\times10^{13}\,L_{\odot}\,$kpc$^{-2}$, and $\kappa_{\rm R,*}=\kappa_{\rm R}(T_*)=2.13$ cm$^{2}$ g$^{-1}$. These values are chosen to be in reasonable agreement with a ULIRG disk.

As shown in KT12, the system with gravity is characterized by two dimensionless numbers, i.e., the dimensionless Eddington ratio
\begin{equation}
f_{\rm E,*} = \frac{\kappa_{\rm R,*} F_*}{g c},
\end{equation}
and the optical depth
\begin{equation}
\tau_* = \Sigma \kappa_{\rm R,*}.
\end{equation}
Physically, the system is initially set to have a temperature of $T_*$ everywhere, $f_{\rm E,*}$ and $\tau_*$ are the initial Eddington and optical depth of the system.

The characteristic sound speed is defined by
\begin{equation}
c_{s,*}^{2}= \frac{k_{\rm B}T_*}{\mu m_{\rm H}}.
\end{equation}
The scale height $h_*$, density $\rho_*$ and time $t_*$ are
\begin{equation}
h_* = \frac{c_{s,*}^{2}}{g},
\qquad
\rho_* = \frac{\Sigma}{h_{*}},
\qquad
t_* = \frac{h_*}{c_{s,*}}.
\end{equation}

The VET and Monte Carlo results imply that a wind can be launched for an initially sub-Eddington system with $f_{\rm E,*}\sim 1$ (i.e. an Eddington factor less than, but near unity), while the FLD and M1 closure methods imply that a dusty wind can be launched only for $f_{\rm E,*}>1$. Regardless, if a dusty wind has been launched by radiation pressure or other mechanisms, it has already overcome its gravitational potential at the base of the system. Following KT13, we focus on the limit of $g\rightarrow 0$, i.e., $f_{\rm E,*}\rightarrow \infty$ in our simulations. In this case an accelerating wind without gravity gives an upper limit of momentum transfer between radiation and dusty gas. 

For this we need to define another set of natural units for the gravity-free system. We use a characteristic acceleration
\begin{equation}
g_{a}=\frac{\kappa_{\rm R,*}F_0}{c},
\end{equation} 
which parameterizes the radiation force on the dust. The units of length, time and density are defined using $g_{a}$ instead of $g$:
\begin{eqnarray}
&&h_{a}=\frac{c_{s,*}^{2}}{g_{a}}=\frac{h_*}{f_{\rm E,*}}
\qquad
t_{a}=\frac{h_{a}}{c_{s,*}}=\frac{t_{*}}{f_{\rm E,*}}\nonumber\\
&&\rho_{a}=\frac{\Sigma}{h_a}=f_{\rm E,*}\rho_*.
\end{eqnarray}
Note that the definitions of $h_a$, $t_a$ and $\rho_a$ are different from KT13. In KT13, $h_a$, $t_a$ and $\rho_a$ are functions of $\tau_*$, but we set these variables to be independent on $\tau_*$, which provides common time and length scales for any choice of $f_{E,*}$.

\subsection{Initial Conditions}

Since the gravitational field is set as $g=0$ ($f_{\rm E,*}\rightarrow \infty$), the simulation results only depends on $\tau_*$. In this paper, we run four 2-dimensional simulations in the $(x,y)$ plane with three values of $\tau_*$. Two types of boundary conditions -- hydrodynamic boundary and radiation boundary are set up for all simulations. Periodic boundary conditions are imposed in the horizontal direction ($x-$direction) on both the radiation and hydrodynamic variables. Reflecting and outflow boundary conditions are used on hydrodynamic variables at the bottom and the top of the vertical direction ($y-$direction) respectively, and inflow and vacuum boundary conditions are setup at the bottom and the top respectively.

Tables \ref{tab_parameter1} summarizes simulation parameters for our runs. T3H and T3L correspond to $\tau_*=3$, T10 correspond to $\tau_*=10$, and T1 corresponds to $\tau_*=1$. We run T3H and T10 with a high resolution $\Delta x/h_a=0.25$, and T1 and T3L with a low resolution $\Delta x/h_a=0.512$. In D14 (see also \citealt{Rosdahl15} and \citealt{Tsang15}), the isothermal dusty atmosphere is initialized with initial density perturbation, which seeds the growth of RTI and turbulence. It is reasonable to assume that a wind launched at the base of a galaxy have already been in a fully turbulent state with small-scale structures. The initial conditions for gas-wind interaction in KT13 are chosen from the end states of the simulations from KT12. Similarly, in this paper we choose the initial conditions from the end states in D14. 

%For all simulations, reflecting boundary conditions are injected at the bottom of the domain box $y=0$. On the other hand, periodic boundary conditions in the $x$ direction are imposed by iterating the short characteristics solution to convergence. 

\begin{table}
\begin{center} 
Simulation Parameters

\begin{tabular} {lcccc}
\hline\hline
Run & IC & $[L_x \times L_y]/h_a$ & $N_x \times N_y$ & $\Delta x/h_a$\\
\hline
T3H   & T3\_F0.5 & $256\times 4096$ & $1024 \times 16384$ & 0.25\\
T10    & T10\_F0.5 & $256\times 4096$  & $1024 \times 16384$ & 0.25\\
T1     & T1\_F0.5L  & $256\times 8192$  & $500 \times 16000$ & 0.512 \\
T3L   & T3\_F0.5L & $256\times 8192$ & $500 \times 16000$ & 0.512 \\
\hline \hline
\end{tabular}
\end{center}
\caption{The initial conditions (IC) show the corresponding runs with gravity in D14. $L_x \times L_y$ is the size of the computational box in unit of $h_a$. $N_x \times N_y$ is the zones in the box. $\Delta x/h_a=0.25$ for the high resolution, and $\Delta x/h_a =0.512$ for the low resolution.}\label{tab_parameter1}
\end{table}

D14 considered various runs with a range of $\tau_*$ and $f_{\rm E,*}$ with gravity. In particular, we focus on three runs: T10\_F0.5 ($\tau_*=10$, $f_{\rm E,*}=0.5$), T3\_F0.5 ($\tau_*=3$, $f_{\rm E,*}=0.5$), and T1\_F0.5 ($\tau_*=1$, $f_{\rm E, *}=0.5$) in D14. The size of the box for three runs are $[L_x \times L_y]/h_*=512\times1024$, with a resolution of $L_{(x,y)}/N_{(x,y)}=0.5 h_*$. We take the simulations results from T3\_F0.5 and T10\_F0.5. The gas in the atmosphere is accelerated by radiation, and eventually reaches the top of the box as unbound outflow. The gas in T3\_F0.5 is accelerated to $\sim 1024 h_*$ at $t \sim 80 t_*$, and turbulence is well developed within the gas. We take the gas at $t = 80 t_*$ as the initial state for run T3H. On the other hand, the gas in T10\_F0.5 reaches the top of the box much earlier than that in T3\_F0.5 due to a higher optical depth and a larger radiation force on the gas. We take the gas at $t = 37.5 t_*$ as the initial state for run T10. Note that the resolution for T3H and T10 are the same as T3\_F0.5 and T10\_F0.5, but the length scale has been changed from $h_*$ to  $h_a$ in this paper.

We also run two simulations T1\_F0.5L and T3\_F0.5L, which have the same setup as T1\_F0.5 and T3\_F0.5 in D14 respectively, but using a lower resolution $L_{(x,y)}/N_{(x,y)}=1.024 h_*$. In constrast to other cases with $\tau_* >1$, the gas in T1\_F0.5 is only accelerated to a maximum height of $\sim 200 h_*$ at $t \sim80 t_*$, then falls back to the base of the system, and eventually reaches a quasi-steady state at $t\sim 125 t_*$. We take the gas at $t = 80 t_*$ when the gas reaches its maximum height as the initial state for T1. T3\_F0.5L gives a very similar result as T3\_F0.5 in D14, but a slightly slower acceleration by radiation because of the lower resolution (see D14 for more discussion on the effects of spatial resolution). We choose the initial state of T3L at $t = 90 t_*$ from T3\_F0.5L. Note that T1 and T3L have the same resolution as T1\_F0.5L and T3\_F0.5L.

In all simulations, we expand the vertical direction of the domain box and initialize zones which are beyond the simulation domains in D14 by giving a uniform temperature $T_* = 82\,$K, and $\rho =10^{-10} \rho_*$ as the background. Low resolution runs (T1 and T3LR) were carried out on the \textit{Rivanna} cluster at the University of Virginia, and high resolution runs (T3HR and T10) were carried out on the TACC cluster Stampede.

%However, we slightly change the resolutions and the size of the box and re-run the problems in D14. ... T1 and T3LR are different from those in D14, we first start to run the simulations T1F0.5 and T3F0.5H in the box of .  until the top of the cloud reaches $\sim 512 h_a$. ... Gas in the simulation T1F0.5 is bound, thus we take the ... when the cloud reaches its highest place. 

\begin{figure}[t]
\begin{center}
\includegraphics[width=7.6cm]{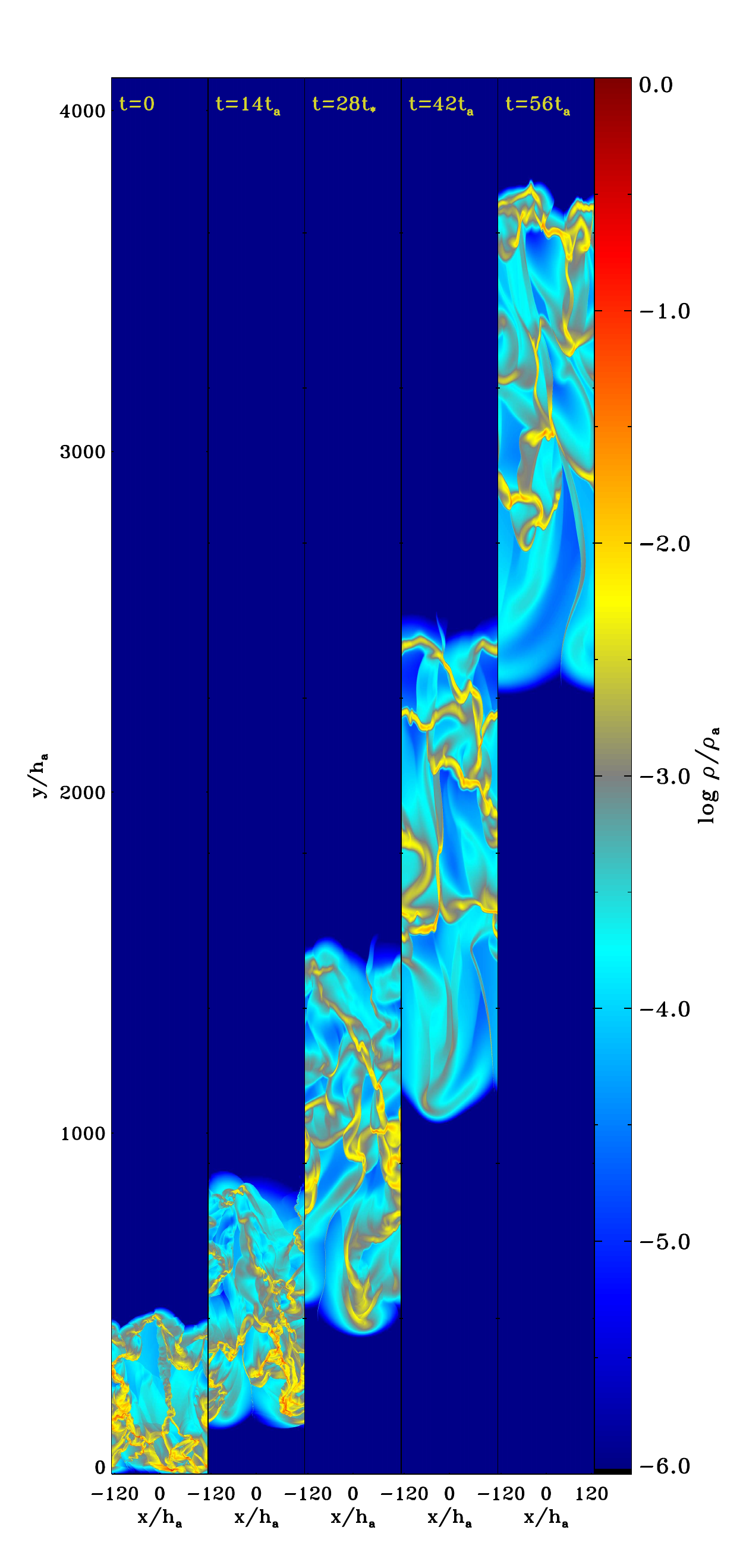}
\end{center}
\caption{Density distribution $\rho$ for five snapshots from run T3H.}\label{T3H}
\end{figure}

\begin{figure}[t]
\begin{center}
\includegraphics[width=7.6cm]{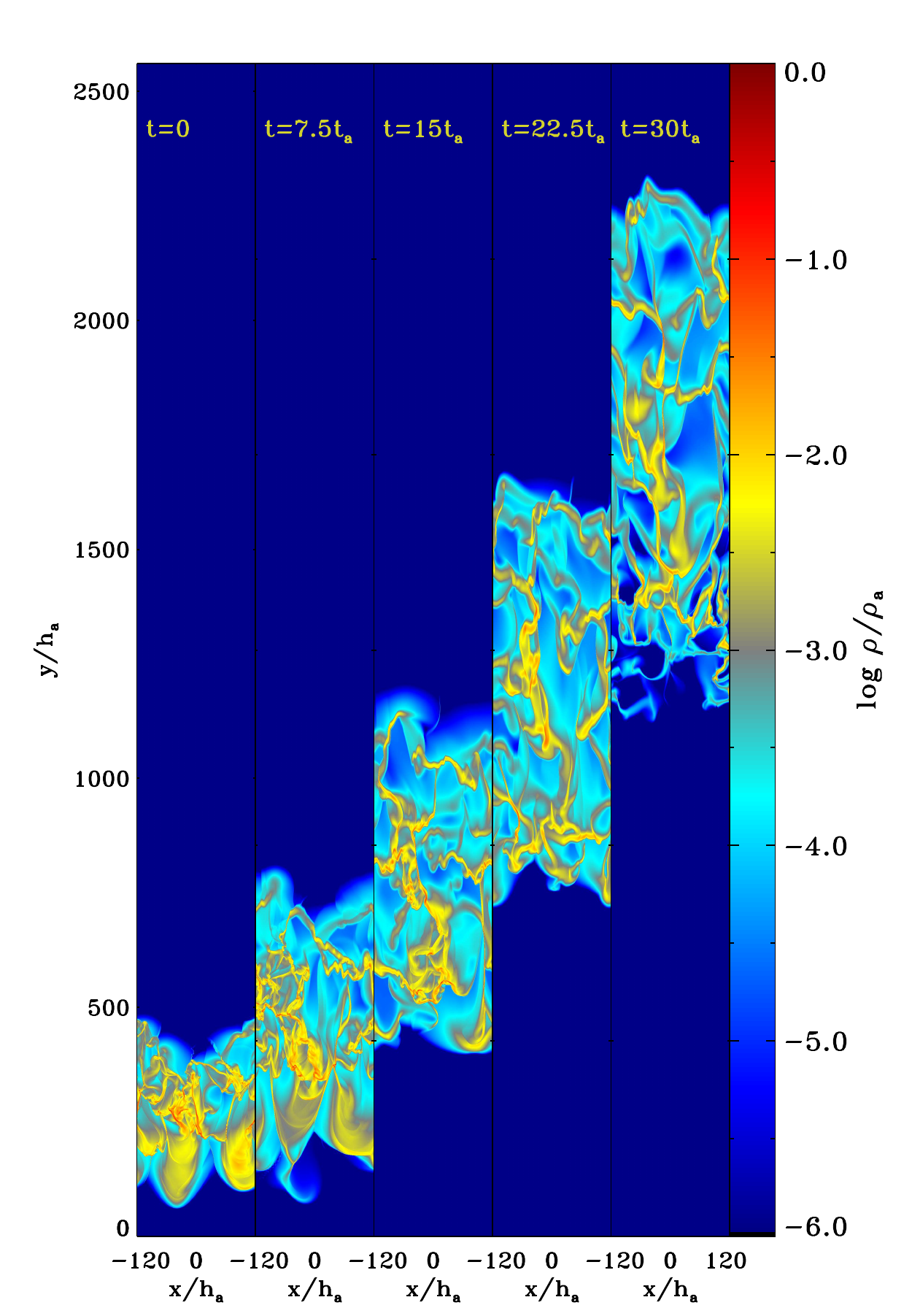}
\end{center}
\caption{Density distribution $\rho$ for five snapshots from run T10.}\label{T10}
\end{figure}

\begin{figure}[t]
\begin{center}
\includegraphics[width=7.6cm]{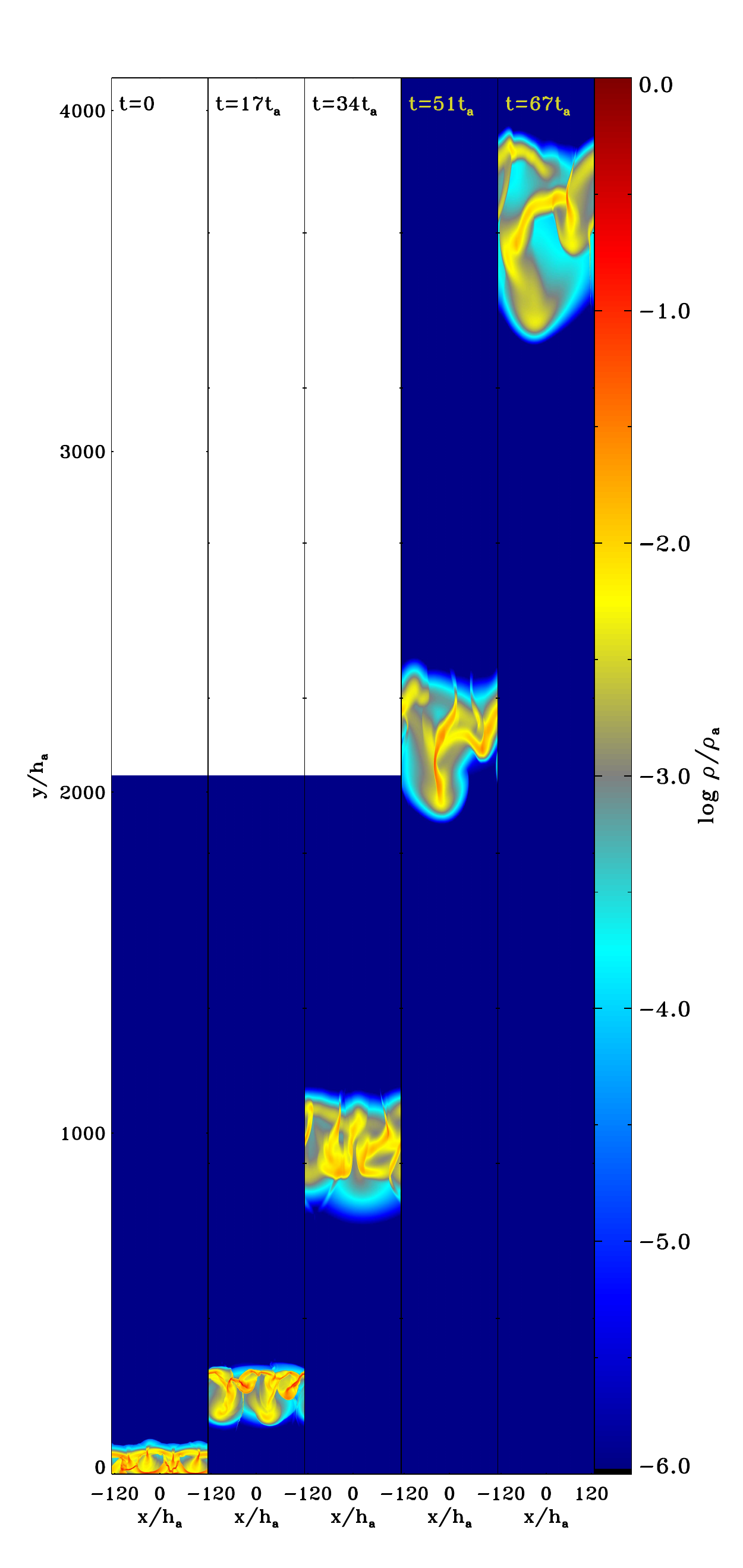}
\end{center}
\caption{Density distribution $\rho$ for five snapshots from run T1.}\label{T1}
\end{figure}

\begin{figure*}[t]
\centerline{
\includegraphics[width=9.5cm]{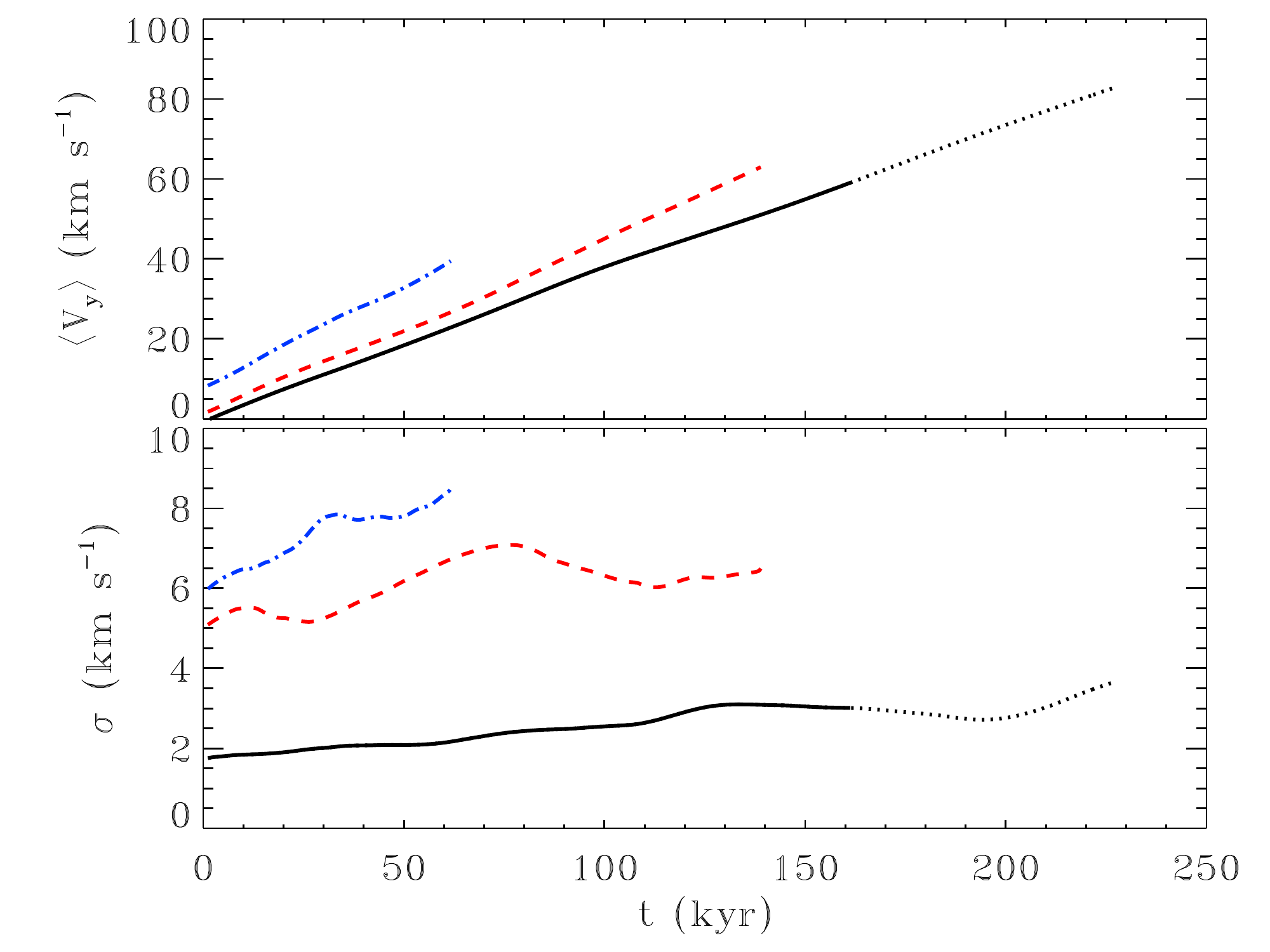}\includegraphics[width=9.5cm]{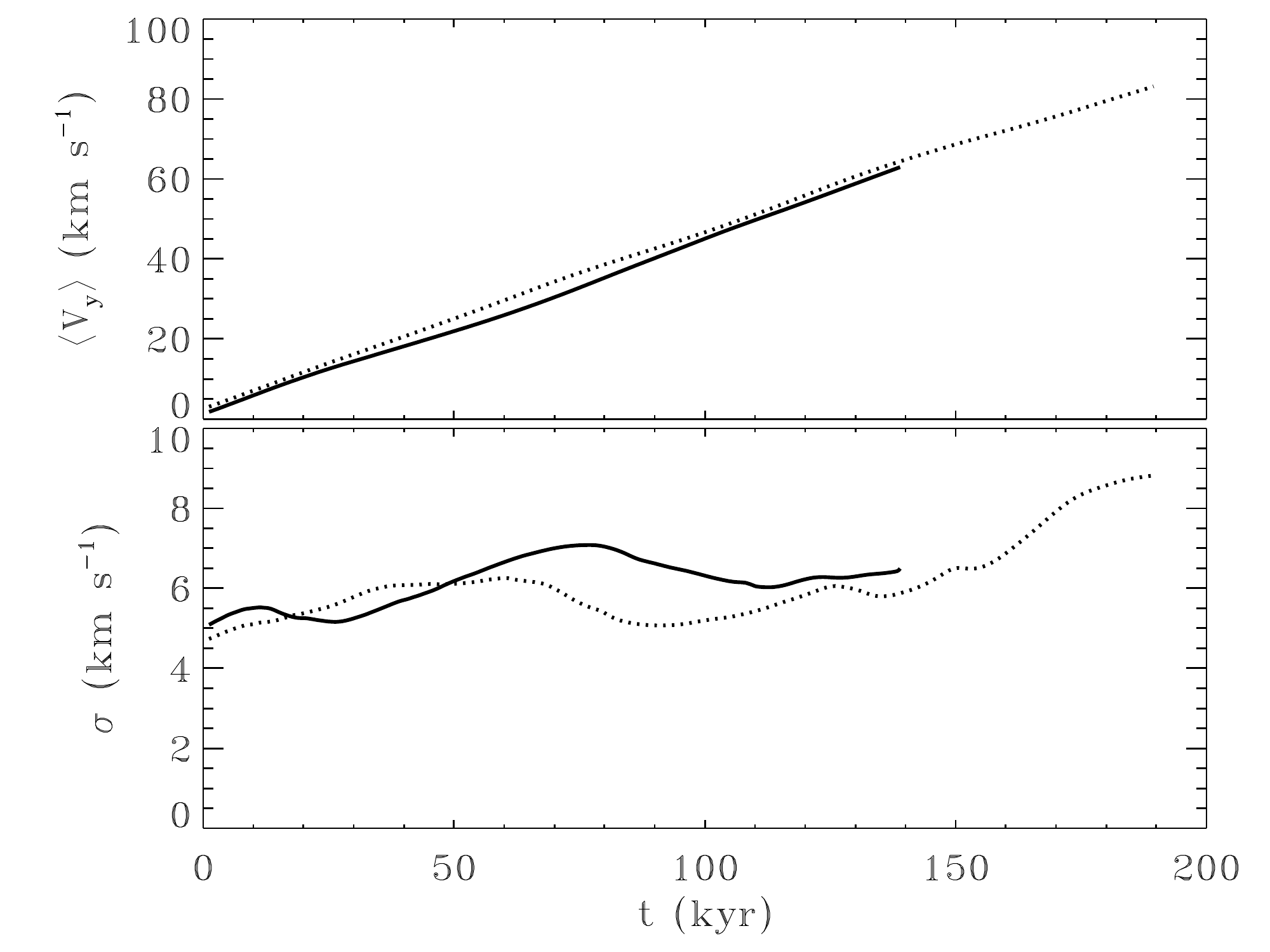}}
\caption{Left: mean gas velocity as a function of time for three runs T1 (black), T3H (red dash) and T10 (blue dash-dotted). The solid black shows the gas acceleration in T1 in a box with the same height ($4096 h_a$) as in T3H and T10, the dotted black shows the gas acceleration in T1 to a height of $8192 h_a$. Right: mean gas velocity as a function of time for T3H (solid) and T3L (dotted). The gas in T3L is accelerated to a height of $8192 h_a$. }\label{velocity1}
\end{figure*}

\begin{figure}
\begin{center}
\includegraphics[width=7.6cm]{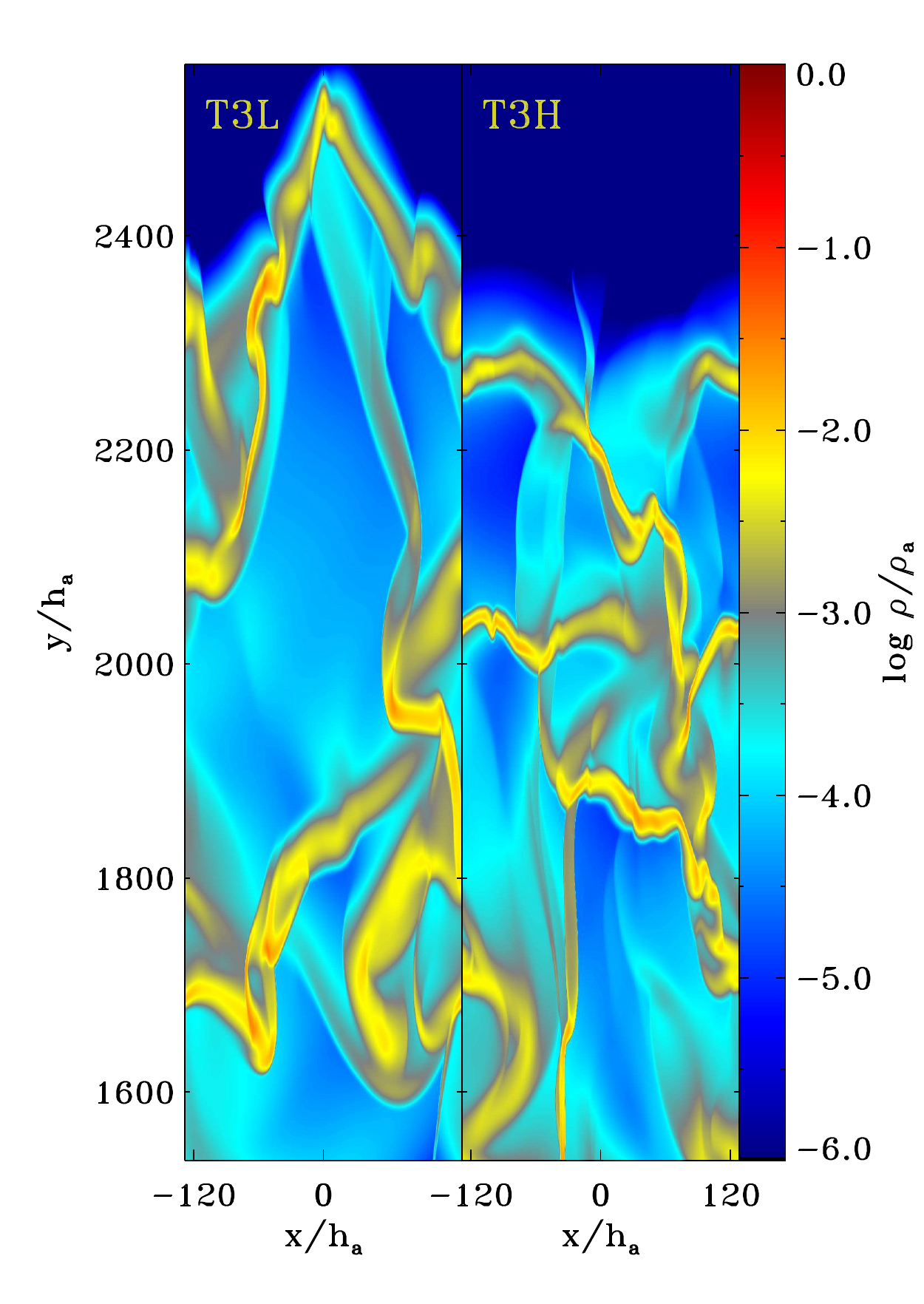}
\end{center}
\caption{Comparison of density distribution $\rho$ between T3L and T3H.}\label{compare2}
\end{figure}

\section{Results}\label{results} 

\subsection{Wind Properties}\label{sec_densityprofile}

We first consider the T3H run. Figure \ref{T3H} shows five snapshots of the density field from this run. Without the gravitational confinement, the gas moves upward and expands in the vertical direction, with the initial filamentary structure stretches out in the radiation field. At $t \sim 59 t_a$, the dense gas hits the upper boundary of the domain, and the gas expands to a thickness of $\sim 1300 h_a$, which covers $\sim 35\%$ of the box. Most of the gas is in a few filaments with $\rho \sim 10^{-3} - 10^{-2} \rho_*$, in between the filaments the volume is filled with a gas of $\rho \gtrsim 10^{-5}\rho_*$. This result is different from KT13 (their Figure 2), in which more extended filamentary structure is driven by radiative RTI, and the vertical extent of the gas eventually becomes comparable to the vertical size of the entire computational box.

Figure \ref{T10} shows density snapshots for the run T10. Due to a higher initial optical depth, the gas is accelerated faster than that in T3. The initial turbulent filamentary structure is stretched along the direction of motion due to the differential acceleration by the radiation field and the gas has a larger velocity dispersion than the gas in T3. Between the filaments of dense gas, the volume has lower density $\rho \lesssim 10^{-6}\rho_*$.  After $t=30\,t_a$, the relative velocities at some shock fronts in the T10 run become very high (Mach numbers of $\lesssim 100$).  In regions where shock fronts cross obliquely, the temperature spikes in low density zones adjacent to the shock front. The algorithm compensates on the following timesteps by generating a large radiation flux that then artificially heats neighboring optically thick zones.  From this point on energy conservation is violated at a few percent level.  By itself, this modest violation of energy conservation might not be troubling, but this uncontrolled heating produces artificially elevated temperatures at interfaces between low density channels and high density filaments. Due to the $T^2$ temperature dependence of the opacity, the radiation forces on the edges of the low density channels causes them to expand, creating voids that are not seen in other simulations or at earlier times in this simulation.  Reducing the timestep reduces the temperature jumps, altering subsequent evolution. However, running for extended time with a significantly lower timestep would be prohibitively computationally expensive so we halt this run at $t=30\,t_a$.

Figure \ref{T1} shows gas acceleration for $\tau_* =1$. Simulations in D14 shows that gas in a gravitational field with $\tau_*\leq 1$ and $f_{E,*} \leq 0.5$ should fall back to the bottom of the system and maintains a quasi-steady state. In the absence of gravity the gas is accelerated and becomes unbound, and we study the behavior of the unbound gas. It reaches the top of the box at $t\sim 68 t_a$ and spreads over a vertical height of $\sim 500 h_a$, smaller and with lower velocity dispersion than the gas in T3 and T10. 
 
Gas acceleration with various optical depths can be clearly seen in the mass-weighted mean velocity, which is given by
\begin{equation}
\langle \mathbf{v} \rangle = \frac{1}{M}\int \rho \mathbf{v} dV,
\end{equation}
and mass-weighted velocity dispersion
\begin{equation}
\sigma^2_{i} = \frac{1}{M} \int \rho (v_i-\langle v_i\rangle)^2 dV,
\end{equation}
respectively, where $M=\int \rho dV$ is the total mass in the atmosphere. The left panel of Figure \ref{velocity1} shows the mean velocity in $y$ direction $\langle \mathbf{v}_y \rangle$, and the total velocity $\sigma=\sqrt{\sigma_x^{2}+\sigma_y^{2}}$ for T1, T3H and T10 runs.  For convenience, we convert the dimensionless units in the simulations to cgs units. The velocity $\langle \mathbf{v}_y \rangle$ increases almost linearly with time. The timescale of gas acceleration to 50 km s$^{-1}$ is $\sim 100\,$kyr, which is comparable to the time to launch a wind from the base of the system. Gas with lower initial optical depth $\tau_*$ receives a slower acceleration. For example, $\langle \mathbf{v}_y \rangle$ reaches 50 km/s at $t\sim 150\,$kyr in T1, but reaches the same velocity with a shorter time $t\sim110\,$kyr in T3H. 
%Dispersion velocity growths with time initially in T1 and T3H, and becomes flat at later times. 
Also, lower $\tau_*$ leads to lower dispersion velocity. Velocity dispersion $\sigma$ in T1 grows from $\sim2$ km/s to 3.5 km/s at $t\sim 200\,$kyr, while $\sigma$ in T3H increases from 5 km/s to 7 km/s at $t\sim 80$\,kyr, then oscillates at $\sim6-7$ km/s at later time.

Note that the velocities obtained by radiation-pressure-acceleration in Figure \ref{velocity1} are far below the observed velocities of cold clouds in nearby starbursts such as M82 (\citealt{Walter02}; \citealt{Leroy15}), NGC 253 (\citealt{Bolatto13}; \citealt{Walter17}), Mrk 231 (\citealt{Rupke11}; \citealt{Gonzalez14}; \citealt{Feruglio15}), or other star-forming galaxies (e.g., \citealt{Heckman00}; \citealt{Veilleux05}; \citealt{Rupke02, Rupke05a, Rupke05b, Rupke05c}; \citealt{Martin05}; \citealt{Weiner09}; \citealt{Chen10}; \citealt{Erb12}; \citealt{Kornei13}), which can reach hundreds or even thousands of km s$^{-1}$. However, in our simulations we only study wind propagation within a vertical height of $\sim 5-10\,$pc. An estimate of the momentum transfer from the radiation to the gas on a larger scale is discussed in Section \ref{sec_discussion}.

\subsection{Spatial Resolution}\label{sec_resolution}

We have performed simulations with the same initial conditions $\tau_*=3$, with a high resolution in the T3H run, and a low resolution in the T3L run. Since the low resolution run is less expensive, we run T3L a bit longer than T3H. We run T3L in a box with a vertical height of $8192 h_a$, which is twice as that in the higher-resolution T3H. The right panel of Figure \ref{velocity1} compares $\langle \mathbf{v}_y \rangle$ and $\sigma$ in T3H and T3L. Although the initial conditions from T3\_F0.5 and T3\_F0.5L are slightly different, the two runs with different initial inputs show very similar acceleration. The velocity dispersion $\sigma$ increases more quickly in T3H at $t\lesssim 80\,$kyr, but both $\sigma$ become flat at $80\,$kyr $\lesssim t \lesssim\,$140 kyr, then $\sigma$ in T3L slightly increases to $9\,$km s$^{-1}$ by the end of the simulation.

%Figure \ref{T3L} shows the snapshots of the density field from T3L. The gas reaches the height of $4096 h_a$ at $t\sim106 t_a$, and hits the top of the box $8196 h_a$ at $t\sim166 t_a$. The right panel of Figure \ref{velocity1} compares $\langle \mathbf{v}_y \rangle$ and $\sigma$ in T3H and T3L. Although the initial conditions from T3\_F0.5 and T3\_F0.5L are slight different, the two runs with different initial inputs show very similar acceleration. $\sigma$ increases more quickly in T3H at $t\lesssim 80\,$kyr, but both $\sigma$ become flat at $80\,$kyr $\lesssim t \lesssim\,$140 kyr, then $\sigma$ in T3L slightly increases to $9\,$km s$^{-1}$ by the end of the simulation. 

Figure \ref{compare2} shows the snapshots of the density distribution $\rho$ from T3L and T3H at a same time $t=40 t_a$. The gas in T3L is accelerated more quickly than that in T3H. The front of the gas in T3L reaches a height of $h\sim2600 h_a$, when the gas in T3H only reaches $h\sim 2300 h_a$. The structure at the top of the gas in two runs are different: T3L shows a slighltly denser front head than the gas in T3H. The similar acceleration in the two cases suggests resolving $h_*$ might not be essential for obtaining the correct value for the bulk radiative acceleration of the outflow.  This should bode well for larger scale simulations of radiative outflows where resolution of $h_*$ would require prohibitively high resolution.  

\begin{figure*}
\begin{center}
\includegraphics[width=9.7cm]{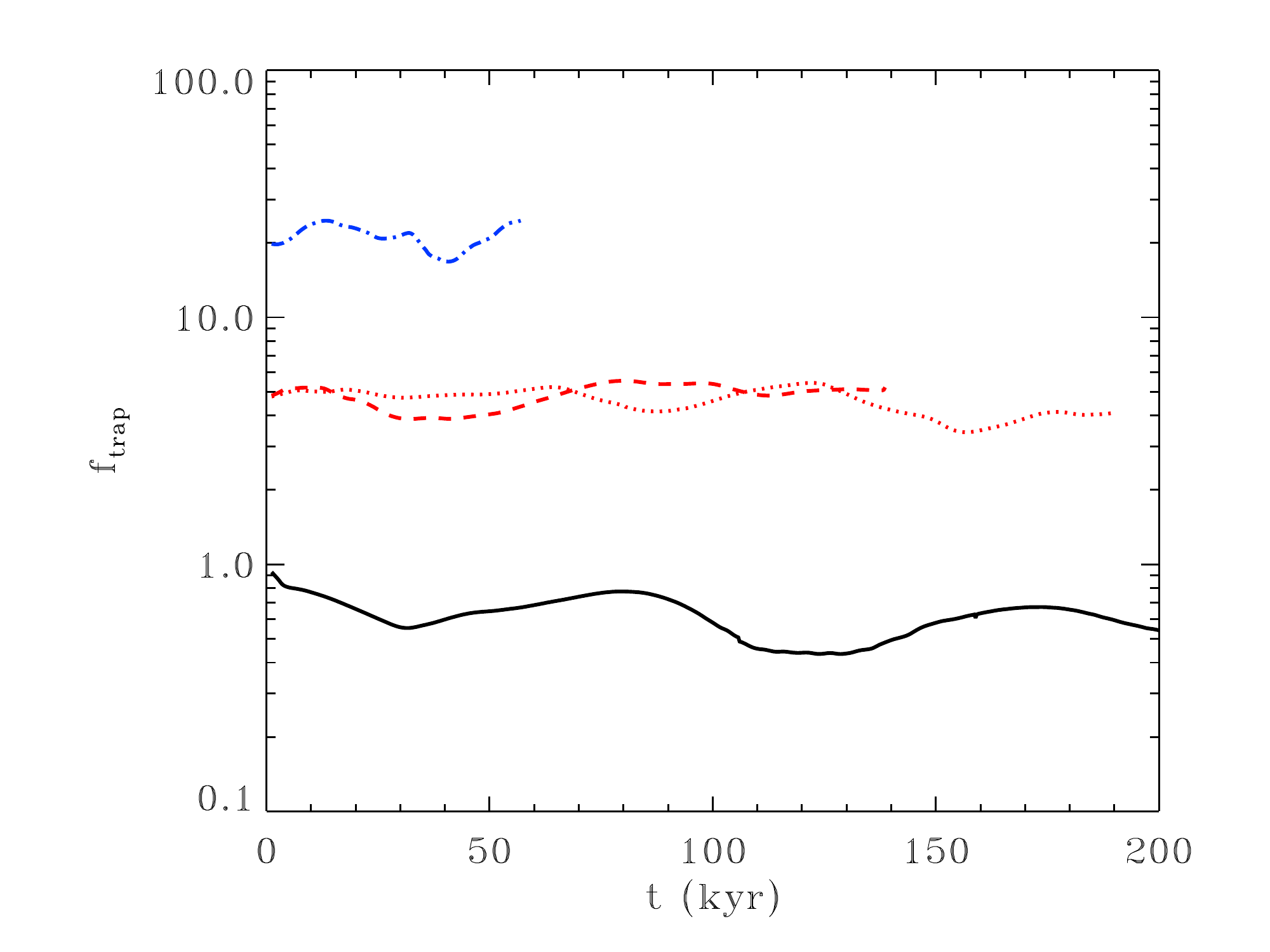}
\end{center}
\caption{Trapping factors as a function of time for four runs: T1 (black solid), T3L (red dotted), T3H (red dashed) and T10 (blue dash-dotted).}\label{trapping}
\end{figure*}

\begin{figure*}
\begin{center}
\includegraphics[width=9.7cm]{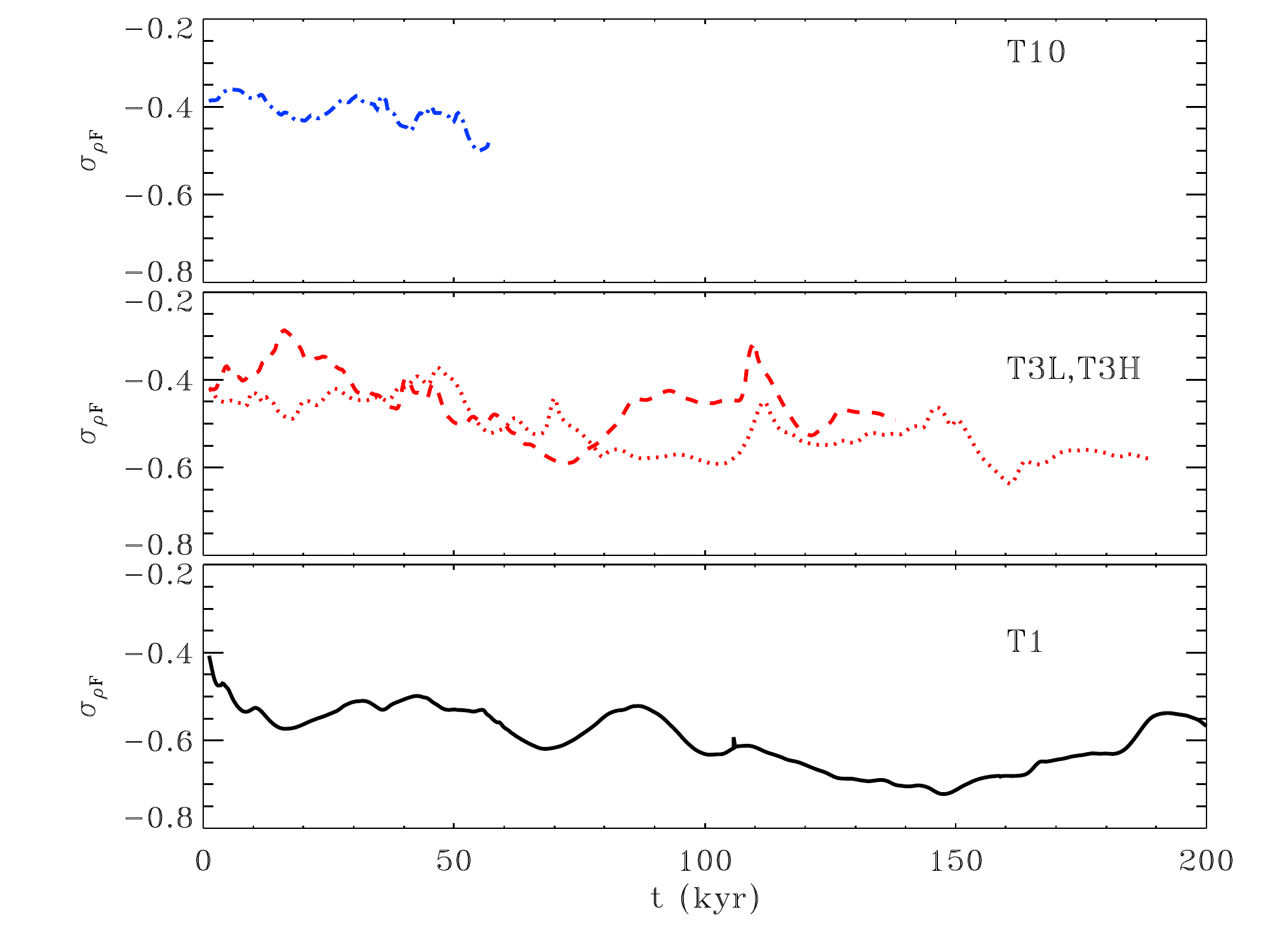}
\end{center}
\caption{Correlation between density and flux $\sigma_{\rho F}$, which is averaged over grid zones with $\rho \geq 10^{-6}\,\rho_{*}$. The lines have the same meaning as in Figure \ref{trapping}.}\label{correlation}
\end{figure*}

\begin{figure*}[t]
\centerline{
\includegraphics[width=12.0cm]{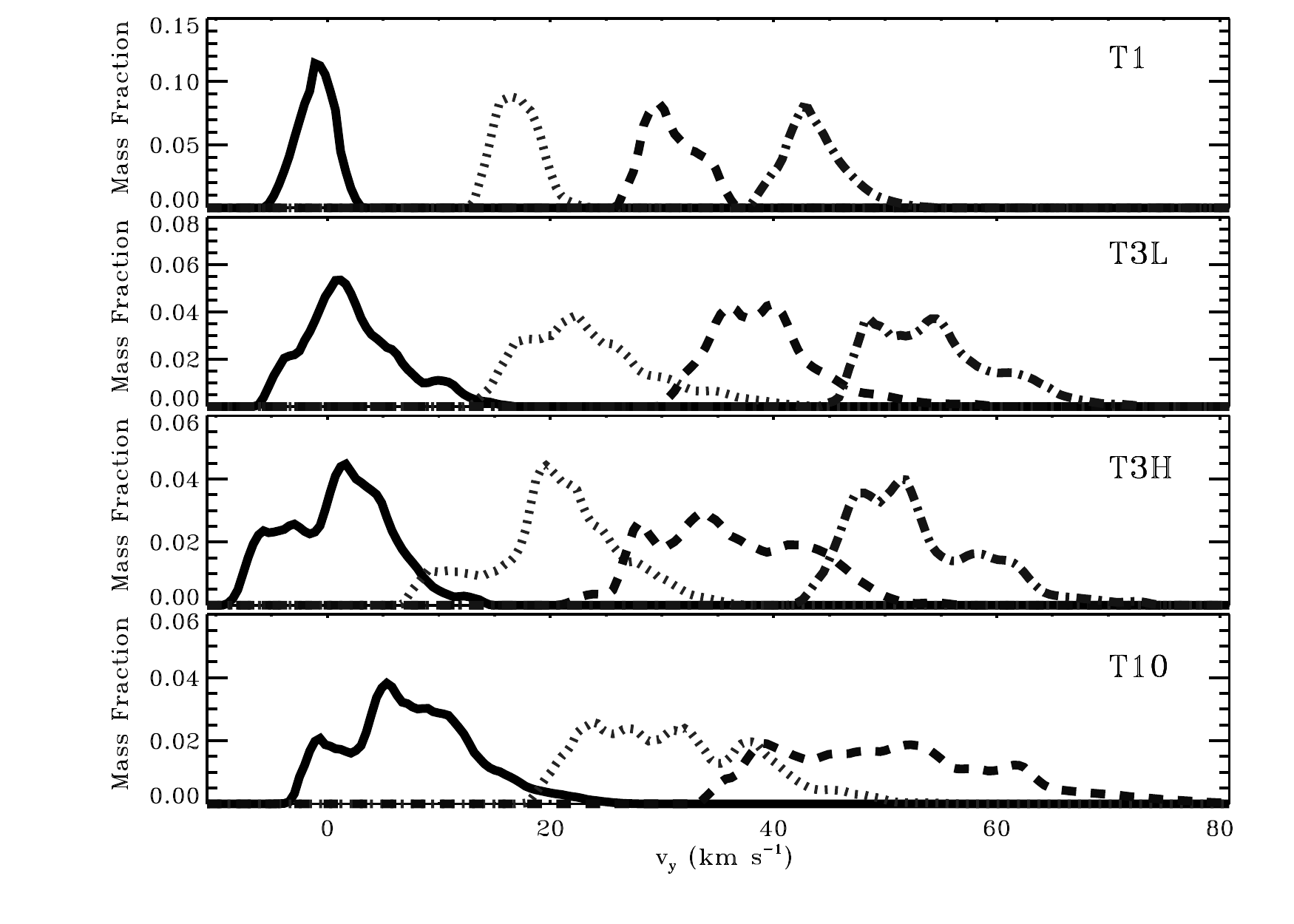}}
\caption{Velocity distribution functions for run T1, T3L, T3H and T10 at $t/t_a=0$ (solid), 15 (dotted), 30 (dashed) and 45 (dash-dotted).}\label{Vdistribution}
\end{figure*}

\begin{figure}[t]
\begin{center}
\includegraphics[width=9.0cm]{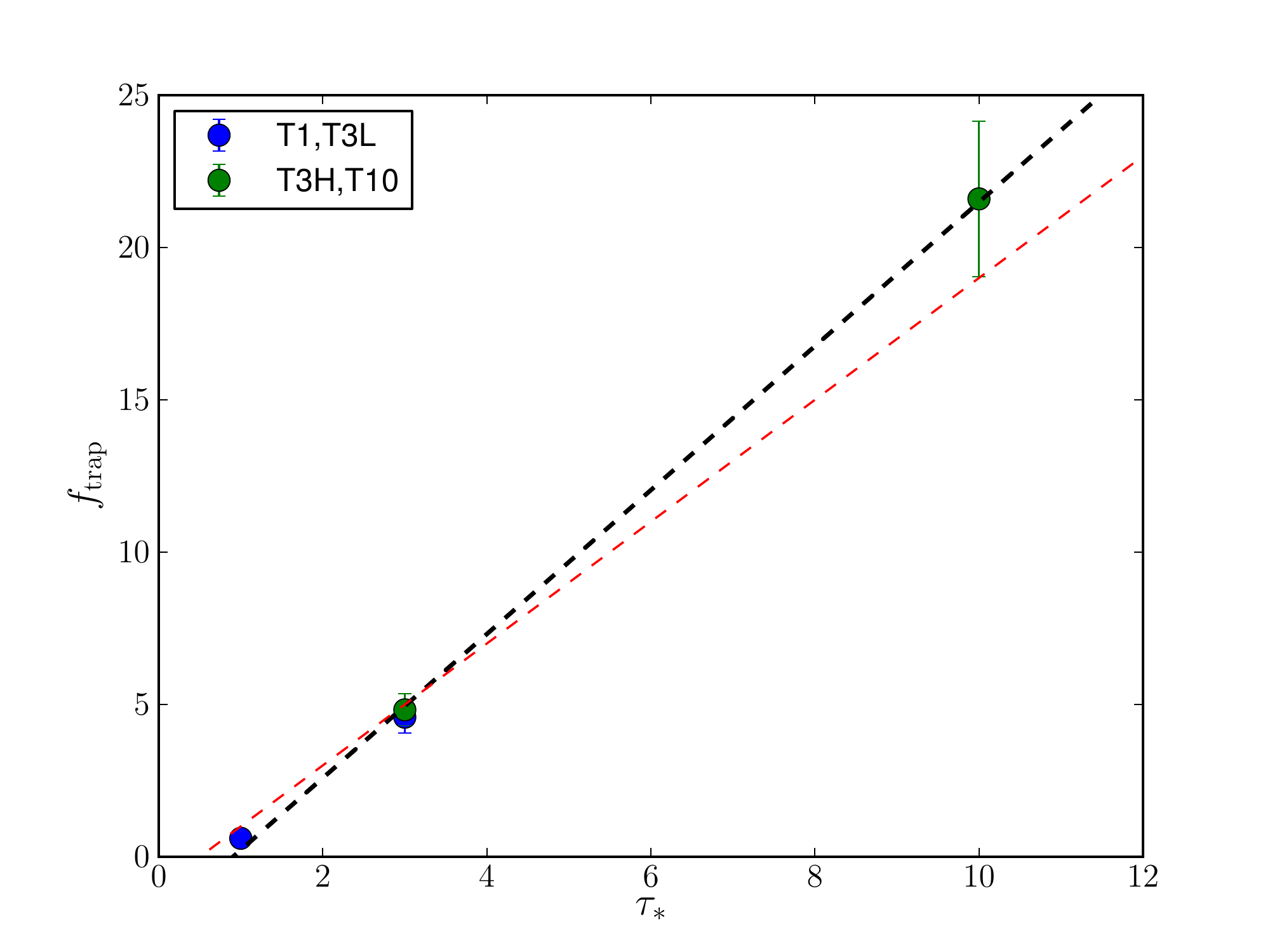}
\end{center}
\caption{Trapping factors as a function of optical depth for four runs. The points shows the time-averaged value, and the error bars are the standard deviation. The black dashed line is the linear fitting of the points, and red dashed line is $f_{\rm trap}=\tau_*/(f_{E,*})_{0}-1$. }\label{fitting}
\end{figure}
\subsection{Trapping Factor}\label{sec_trapping}

We study the momentum coupling between the infrared radiation field and the gas. Without gravity, the y-component momentum equation of the gas is
\begin{equation}
\frac{d \langle v_y \rangle}{d t}=f_{\rm rad},\label{momentum1}
\end{equation}
where $f_{\rm rad}$ is defined as the mean radiation force per unit mass (acceleration)
\begin{equation}
f_{\rm rad}=\frac{1}{c}\frac{\langle \kappa_{\rm R}\rho F_{ry} \rangle}{\langle \rho \rangle}.
\end{equation}
Following KT13, we define the trapping factor $f_{\rm trap}$ in a gravity-free field by
\begin{equation}
1+f_{\rm trap} = \frac{f_{\rm rad}}{f_{\rm rad,dir}}\label{trapping1}
\end{equation}
where $f_{\rm rad,dir}$ is the momentum flux per unit mass of the directly injected radiation field. We have $f_{\rm rad,dir}=F_{ry}/(c\langle \rho \rangle L_y)=F_*/(c\langle \rho \rangle L_y)$, where $L_y$ is the vertical height of the computational domain. Thus, equation (\ref{momentum1}) can be re-written as
\begin{equation}
f_{\rm trap}=\frac{t_a \tau_*}{c_{s,*}}\frac{d \langle v_y \rangle}{d t}-1.\label{momentum2}
\end{equation}

The trapping factor $f_{\rm trap}$ measures the momentum transfer from the radiation to the gas. The upper limit of trapping in analytic models adopts $f_{\rm trap}\sim \tau_{\rm IR}$, where $\tau_{\rm IR}$ is the infrared optical depth of the system. In our simulations, the initial $f_{\rm trap}^{0}$ is obtained from the end state of the gas at the base of the system with gravity (KT12 and D14)
\begin{equation}
1+f_{\rm trap}^{0}=L_y \frac{\langle \kappa_R \rho F_{ry}\rangle}{\langle F_{ry} \rangle}=\frac{f_{\rm rad}\tau_*}{g f_{E,*}}=\frac{\tau_* f_{E,V}}{f_{E,*}},
\end{equation}
where $f_{E,*}$ is the fiducial Eddington ratio in the simulation with gravity ($f_{E,*}=0.5$), and
\begin{equation}
f_{E,V}=\frac{f_{\rm rad}}{g} \label{fEv}
\end{equation}
is the Eddington ratio computed using the initial gravity $g$. According to KT12 and D14, $f_{E,V} \sim 1$ due to the radiative RTI regulation, therefore,
\begin{equation}
f_{\rm trap}^{0} \simeq \frac{\tau_*}{f_{E,*}}-1.\label{ftrap0}
\end{equation} 
We have $f_{\rm trap}^{0} \simeq 1$ for $\tau_*=1$,  $f_{\rm trap}^{0} \simeq 5$ for $\tau_*=3$, and $f_{\rm trap}^{0} \simeq 19$ for $\tau_*=10$.

KT13 found that $f_{\rm trap}$ without gravity significantly deceases from $f_{\rm trap}^{0}$ to a smaller value, which they attributed to the radiative RTI. Figure \ref{trapping} shows trapping factor as a function of time in our simulations. In contrast, we do not see any significant evolution of the trapping factor.  Comparison of $f_{\rm trap}$ for T3L and T3H suggests that the trapping property is insensitive to the resolution. The values of $f_{\rm trap}$ are largely consistent with the values $f_{\rm trap}^0$ inferred from the D14 runs with gravity. Thus, it is perhaps somewhat surprising that the runs performed here with $g=0$ see little evolution of the trapping factor.  One possibility is that the RTI has little to no affect on trapping factor in these runs with $g=0$ and the simulations simply retain knowledge of their initial density and flux distributions. If present, the RTI is expected to largely shape the trapping factor through its effect on the flux -- density relationship, so we 
calculate the correlation between density $\rho$ and the vertical component of radiation flux $F_{ry}$
\begin{equation}
\sigma_{\rho F}=\frac{\langle (\rho-\langle \rho \rangle)(F_{ry}-\langle F_{ry} \rangle)\rangle}{\sqrt{\langle  (\rho-\langle \rho \rangle)^{2}\rangle} \sqrt{\langle (F_{ry}-\langle F_{ry} \rangle)^{2}\rangle}}.
\end{equation}
We compute $\sigma_{\rho F}$ over the whole simulation domain, but only including grid zones with $\rho \geq 10^{-6}\rho_*$.  The density floor in the correlation excludes the background region where the density is low and flux is near the fiducial value.  If this region is not excluded, it skews the correlation due to the large fraction of the simulation volume at these low densities.   

Since density and flux are anti-correlated, we find a negative value for $\sigma_{\rho F}$, with variations on shorter timescales but no long term evolution in any run. We also find that $\sigma_{\rho F}$ is, on average, higher for larger $\tau_*$. The shorter timescale variation of $\sigma_{\rho F}$ with time does not closely track the variation of $f_{\rm trap}$ for any of the runs, suggesting that effects other than the flux -- density correlation are impacting the trapping factor. The behaviors of $f_{\rm trap}$ and $\sigma_{\rho F}$ in the VET simulations are different from that in the FLD simulation (see Appendix \ref{section_FLD}, Figure \ref{fig_FLD}). Similar to KT13, we find that the trapping factor $f_{\rm trap}$ drops with time in the FLD run. This decrease of $f_{\rm trap}$ matches a trend towards increasingly negative (more anti-correlated) $\sigma_{\rho F}$ with time. As in the VET runs, there is no clear correspondence between variations in $\sigma_{\rho F}$ and $f_{\rm trap}$ on shorter timescales, but the overall downward trend is suggestive that the simulation is allowing a larger fraction of the radiation flux to escape through low density channels as the run progresses.

These results suggest that radiative RTI has relatively little impact on the long term evolution of the flux--density correlation or trapping factor in our VET simulations. Analysis of the linear instability to radiative RTI \citep{Jacquet11} in optically thin and adiabatic limits suggests these flows should be linearly stable when $g \rightarrow 0$.  This does not preclude non-linear interactions that cause channels to develop or widen, but neither is there a clear motivation for radiative RTI to have a strong impact on the density structure and resulting momentum coupling in this limit where $g \rightarrow 0$.

%However, we still find that $f_{\rm trap} \sim f_{\rm  trap}^0$ in all the runs, which can be interpreted simply as the solution preserving its initial condition.  Comparison of our results with KT13 shows that their simulations form long, vertically oriented low-density channels where flux can escape, while our simulations do not. The analysis from D14 suggests that FLD acts to more strongly reinforce the development of non-linear structure and this same tendency may be at play here. Analysis of the linear instability to radiative RTI \citep{Jacquet11} in optically thin and adiabatic limits suggests these flows should be linearly stable when $g \rightarrow 0$.  This does not preclude non-linear interactions that cause channels to develop or widen, but neither is there a clear motivation for radiative RTI to continue to shape the density structure in a manner that would have strong impact on the momentum coupling between gas and radiation.

\subsection{Velocity Distribution}

Figure \ref{Vdistribution} shows mass-weighted velocity probability distribution functions (PDFs) in y-direction for four runs: T1, T3L, T3H and T10. Since the initial condition for T1 is quasi-steady, the velocity distribution is nearly symmetric at $v_y=0$. On the other hand, the initial velocity distributions for T3L, T3H and T10 are asymmetric with a tail extending to $v_y\sim 20-30$ km s$^{-1}$, indicating that most of the gas has already been accelerated at the base of the system. As time evolves, the PDFs for all runs shift to higher velocity of $v_y$. Higher $\tau_*$ gives a higher acceleration, and a larger spreading of velocities. This is consistent with Figure \ref{velocity1} that larger $\tau_*$ leads to faster acceleration and larger velocity dispersion. Also,the panels for T3L and T3H show that resolution does not change the PDF qualitatively.

\section{Discussion}\label{sec_discussion}

\subsection{Momentum Transfer Between Radiation and Gas}

In KT13, a linear fit of $f_{\rm trap}$ is given by $f_{\rm trap}\approx 0.5$ in the limit of $f_{E,*}\rightarrow \infty$. They adopt an interpolation for $f_{\rm trap}$ as a function of $\tau_*$ and $f_{E,*}$, and conclude that winds can only be produced from systems with $f_{E,*}\gtrsim 1$ (super-Eddington limit). However, using the VET method we find different conclusions. In this section we put gravity back to estimate wind acceleration by radiation in a gravity field. 

Note that Figure \ref{trapping} shows that $f_{\rm trap}$ is approximately flat without gravity $f_{E,*}$. Recall the relation that
\begin{equation}
f_{E,V}=(1+f_{\rm trap})\frac{f_{E,*}}{\tau_*}\label{fEV_relation}.
\end{equation}
For $f_{E,*}\lesssim 1$ (but not $f_{E,*}\ll 1$), the radiative RTI regulates equilibrium between infrared radiation and gravity. Thus, we have $f_{E,V}\sim 1$, and $f_{\rm trap}\simeq \tau_*/f_{E,*}-1$. On the other hand, Figure \ref{fitting} shows the time-averaged values of $f_{\rm trap}$ as a function of $\tau_*$ in the limit of $f_{E,*}\rightarrow \infty$ and the linear fitting of the points. We find that the estimate $f_{\rm trap}\sim f_{\rm trap}^{0}\simeq \tau_*/(f_{E,*})_{0}-1$ holds, where $(f_{E,*})_{0}$ is the initial Eddington ratio where the wind is launched.

Including gravity, equation (\ref{momentum1}) can be written as
\begin{equation}
\frac{d \langle v_y \rangle}{d t}=f_{\rm rad}-g\label{dvdt}.
\end{equation}
Combining equations (\ref{trapping1}), (\ref{fEv}), (\ref{fEV_relation}) and (\ref{dvdt}) yields the equation for the net rate of momentum coupling 
\begin{eqnarray}
\frac{d p_{\rm wind}}{d t}&=&(1+f_{\rm trap})\left(1-\frac{1}{f_{E,V}}\right)\frac{L}{c}\label{dvdt2}.
\end{eqnarray}
Here $d p_{\rm wind}/dt$ is the momentum injection as a combination of both infrared radiation acceleration and gravitational deceleration. Note that $f_{E,V}/f_{E,*}  \sim (f_{E,V})_{0}/(f_{E,*})_{0}$, or $f_{E,V}\simeq f_{E,*}(f_{E,V})_{0}/(f_{E,*})_{0}$, and $1+f_{\rm trap} \simeq 1+(f_{\rm trap})_0 = \tau_* (f_{E,V})_0 /(f_{E,*})_0$ according to our simulations, and since $(f_{E,V})_{0}\gtrsim 1$, from equation (\ref{dvdt2}) we obtain
\begin{equation}
\frac{d p_{\rm wind}}{d t} \simeq \frac{\tau_*}{(f_{E,*})_{0}}\left[1-\frac{(f_{E,*})_{0}}{f_{E,*}}\right]\frac{L}{c},
\end{equation}
For an infrared optically thick disk, the gravitational force initially drops faster than the radiation force with the height, $f_{E,*}$ increases monotonically with the height above the disk, the Eddington ratio above the disk is higher than that at the base of the system, i.e. $f_{E,*}>(f_{E,*})_{0}$ (e.g., \citealt{Zhang12}). If we include the direct radiation from the stellar UV light, we have the total momentum injection from radiation
\begin{equation}
\frac{d p_{\rm wind}}{d t} \sim \left[1+\frac{\tau_*}{(f_{E,*})_{0}}\right]\frac{L}{c},
\end{equation}
This is consistent with the result obtained at the base of the system (see Section 5.4 in D14). According to D14, $\tau_*/(f_{E,*})_{0}$ represents the effective infrared optical depth for momentum transfer, which is slightly lower that $\tau_{\rm IR}$ of the system. Here $\tau_{\rm IR}$ can be estimated by the volume-weighted mean optical depth
\begin{equation}
\tau_V = L_{y}\langle \kappa_R \rho \rangle.
\end{equation}
We average $\tau_V$ in our simulations and find $\langle \tau_V \rangle=1.8$ in T1, 7.9 in T3L, 8.5 in T3H and 48.3 in T10, thus, we can define the efficiency $\eta$ where $\eta\tau_{\rm IR}$ is equivalent to $\tau_*/ (f_{E,*})_{0}$. Thus, we find $\eta=0.90$ in T1, 0.71 in T3L, 0.69 in T3H, and 0.47 in T10. Therefore, we conclude that for $f_{E,*}> (f_{E,*})_{0}$, radiation pressure on dust is able to drive an unbound wind. The momentum transfer from the radiation field to the gas is amplified by a factor of $\eta \tau_{\rm IR}$ with $\eta\sim0.5-0.9$, increasing with the optical depth in the atmosphere. 

%On the other hand, if $f_{E,*}< (f_{E,*})_{0}$, the wind may eventually fall back and be bound.  

%\subsection{For winds}

\subsection{Rapidly Star-Forming Galaxies and Starbursts}

Since $\tau_*$ and $f_{E,*}$ are the most important parameters in the simulations, it is worthwhile to estimate them in real rapidly star-forming galaxies and starbursts. KT13 calculated $\tau_*$ and $f_{E,*}$ analytically using a mass-to-light ratio motivated by the starburst99 model (\citealt{Leitherer99}). We estimate $\tau_*$ and $f_{E,*}$ using recent observation data. We consider the gas surface density in a galactic disk is $\Sigma_{\rm g} = 10^{4}\Sigma_{\rm g,4}\,M_\odot$ pc$^{-2}$, the infrared flux is $F_{\rm IR}=10^{13}F_{\rm IR,13}\,L_{\odot}$ kpc$^{-2}$, where $10^{4}\,M_{\odot}$ pc$^{-2}$ (2.1 g cm$^{-2}$), and $10^{13}\,L_{\odot}$ kpc$^{-2}$ are the typical surface densities and fluxes in LIRGs/ULIRGs (e.g., \citealt{Thompson05}). The characteristic temperature in the atmosphere is given by $T_* = (F_{\rm IR}/a_{r}c)^{1/4}$, and the surface gravitational force is $g=2\pi G \Sigma_{\rm g}f_{g}^{-1}$, where $f_{g}=0.5f_{g,0.5}$ is the mass fraction of the gas. Thus, we have
\begin{eqnarray}
&&\tau_*^{\rm max}= 2.8\,\Sigma_{\rm g,4} F_{\rm IR,13}^{1/2},\label{obs1}\\
&&f_{E,*}=0.10 f_{\rm g,0.5} F_{\rm IR,13}^{3/2}(\Sigma_{\rm g,4})^{-1}\label{obs2}.
\end{eqnarray}
Here, equation (\ref{obs1}) gives an upper bound of the infrared optical depth in the atmosphere of the galaxy. We estimate these two values using the most recent observation of LIRGs and ULIRGs measured and compiled by Barcos-Mu\~{n}oz et al. (2016, submitted). They have observed 22 local LIRGs and ULIRGs using the Very Large Array radio observation. We take LIRGS/ULIRGs in their work as a sample. Since molecular gas is presumably the dominant component in LIRGs and ULIRGs, especially in the central regions, we use their molecular gas density $\Sigma_{\rm mol}$ as our estiamte for $\Sigma_g$\footnote{Note that the measurements of $\Sigma_{\rm mol}$ are uncertain, depending on the assumed conversion factor of CO to H$_2$, and the assumption that the emitting area is well-charaterized by the 33 GHz emission. More discussion on these assumptions is given in Barcos-Mu\~{n}oz et al. (2016).}.

Using the data in Barcos-Mu\~{n}oz et al. (2016), we find that most LIRGs/ULIRGs have $f_{E,*}<1$, and about one fourth of them have $f_{E,*}\sim 0.1-1$. The values of $\tau_*^{\rm max}$ are typically large $\tau_*^{\rm max} \gtrsim 1-10$, suggesting a large $\tau_* \gtrsim 1$ in the atmosphere is possible. For example, Arp 220 has molecular gas density $\Sigma_{\rm mol}\sim4.9\times10^{4}\,M_{\odot}$ pc$^{-2}$ and a flux of $F_{\rm IR}\sim 6.1\times 10^{13}\,L_{\odot}$ kpc$^{-2}$, corresponding to $f_{E,*}\sim 0.3$ and $\tau_* ^{\rm max}\sim 30$. Although $\tau_*$ of the atmosphere is much less the total $\tau_*$, we still expect $\tau_* \gtrsim 1$. Our simulation suggests that the infrared radiation may launch dusty gas out of the galaxy, as $f_{E,*}$ drops above the galactic disk, the gas may be accelerated and become unbound. Moreover, an extreme case is given by the ULIRG Mrk 231 (UGC 08058) with $\Sigma_{\rm mol}\sim 1.7\times 10^{5}\,M_{\odot}$ pc$^{-2}$ and $F_{\rm IR}\sim 2.6\times 10^{14}\,L_{\odot}$ kpc$^{-2}$, corresponding to $f_{E,*} \sim 0.8$ and $\tau_{*}^{\rm max}\sim 2.3\times 10^{2}$. Although Mrk 231 has active galactic nucleus activities (e.g., \citealt{Rupke13}), these results suggest that infrared radiation alone could drive a powerful dusty wind in Mrk 231. 

An important caveat to the above analysis is that equations (\ref{obs1}) and (\ref{obs2}) implicitly assume $\kappa_{\rm R} \propto T^2$.  This relation approximately holds only for $T \lesssim 150$K and can lead to a possible overestimation of $f_{E,*}$ and $\tau_*$ for higher temperatures.  For example, the value of $f_{E,*} \sim 0.8$ comes about because $T_* \simeq 150$K and yields $\kappa_{\rm R,*} \simeq 6.8 \rm \, cm^2/g$.  Becoming super-Eddington requires $\kappa_{\rm R}$ to increase  to $8.6 \rm \, cm^2/g$.  It is unclear whether such high infrared dust opacities are obtained in these systems. These values depend on both the dust distribution and dust to gas ratio.  One could alternatively follow \citet{Skinner2015} and formulate a bound on the Eddington ratio in terms of the light-to-mass ratio and an assumed maximum opacity.  Their equation (12) says that super-Eddington fluxes require
\begin{equation}
\kappa_{R} > 15 {\, \textrm{cm}^2 \, \textrm{g}^{-1}}\left(\frac{\Psi}{1700 \ \rm erg \ s^{-1} \ g^{-1}}\right)^{-1},
\end{equation}
where $\Psi$ is the light-to-mass ratio. These results lead one to conclude that both a very large light-to-mass ratio and a high maximum dust opacity are required for radiative pressure alone to drive outflows. 

In general, we find many of the systems in the Barcos-Munoz sample have $f_{E,*} \lesssim 1$ and $\tau_* >1$. Since gas can be launched by radiation from an initially marginally sub-Eddington system, and as $f_{E,*}$ increases with the height above the ULIRG disk, gas may potentially be accelerated to the observed velocities.  If the dust opacities and light-to-mass ratios are sufficiently large, radiation may be able to play a dominant role in driving outflows, but this is most likely only the case in a subset of the most extreme star-forming galaxies.  Radiation pressure may also operate in concert with other driving mechanisms (e.g. supernova, cosmic rays) in less extreme systems.  The key result of our analysis is that it does not seem that RTI alone fundamentally prevents radiative acceleration of outflows.

\section{Conclusions}\label{conclusions}

%We apply the variable Eddington tensor (VET) algorithm to study radiation-pressure-driven dusty winds from optically thick system. 

We study the dusty winds driven by radiation pressure in the atmospheres of rapidly star-forming environments. Krumholz \& Thompson (2013) (KT13) used flux-limited diffusion algorithm to a two-dimensional problem modeling the radiation hydrodynamics (RHD) of a column of gas that is accelerated by a constant infrared radiation flux. We apply the more sophisticated variable Eddington tensor (VET) algorithm to re-examine the problem in KT13. In the absence of gravity, the system, which is characterized by the initial optical depth ($\tau_*$) of the gas and the initial conditions, gives an upper limit on momentum transfer between radiation and gas. We carry out four runs with different $\tau_*$ and varying resolutions. In each simulation, the initial state of the gas is given by the end state of simulation in D14 with the same $\tau_*$ and resolution, but including gravity ($f_{E,*}=0.5$). In D14 the gas evolves only at the base of the system. We expand the vertical direction of the computational domain, and study the wind-gas interaction and momentum coupling between the radiation field and the gas.

We find that the gas spreads out along the height of box with increased mean velocity and velocity dispersion, due to the interactive of the dusty gas and the radiation force. However, the radiative RTI does not seem to be limiting momentum transfer as in KT13. We find that the momentum coupling between gas and radiation in the absence of gravity is similar to that with gravity. The trapping factor $f_{\rm trap}$, which measures the momentum transfer from the radiation to the gas (see equations [\ref{trapping1}]), has the same value to within a factor of two or less at the base of the system. Combing the results in D14, we conclude that dusty gas can be accelerated by radiation even in an initially sub-Eddington system $f_{E,*}<1$, and the momentum from the radiation couples well with the gas during the wind propagation. For $f_{E,*}$ increasing along the height of the system, the momentum transfer from radiation to gas is approximate
\begin{equation}
\frac{d p_{\rm wind}}{d t} \simeq \left\{1+\eta \tau_{\rm IR}\left[1-\frac{(f_{E,*})_{0}}{f_{E,*}}\right]\right\}\frac{L}{c},
\end{equation}
where $(f_{E,*})_{0}$ is the Eddington ratio at the base of the system, $\tau_{\rm IR}$ is the integrated infrared optical depth through the dusty gas, and the efficiency $\eta$ is estimate to be $\sim0.5-0.9$ from $\tau_*=1$ to $\tau_*=10$. Thus, the momentum transfer from the radiation to the wind is not merely $\sim L/c$, but is amplified by a factor of $\eta \tau_{\rm IR}$. Therefore, we conclude that radiation pressure may still be a important mechanism to drive winds in rapidly star-forming galaxies and starbursts.

\acknowledgments

We thank the anonymous referee for helpful comments that improved this manuscript.
We thank Yan-Fei Jiang for helpful discussions and technical assistance. We thank James Stone, Eve Ostriker, Norm Murray, and Loreto Barcos-Mu\~{n}oz for stimulating discussions and/or detailed comments.  D. Z. also thanks Todd Thompson, Mark Krumholz, Evan Scannapieco, Chris Hayward, Nahum Arav, Mike McCourt, Eliot Quataert, Renyue Cen, Mordecai-Mark Mac Low, Greg Bryan, Kohei Inayoshi, Yong Zheng, Zhi-Yun Li, Alberto Bolatto, Sylvain Veilleux, Francesco Tombesi, and Karen Yang for helpful discussions. This work used the Extreme Science and Engineering Discovery Environment (XSEDE), which is supported by National Science Foundation (NSF) grant No. ACI-1053575. This work also used the computational resources provided by the Advanced Research Computing Services (ARCS) at the University of Virginia. S. W. D. acknowledges support from NSF grant AST-1616171 ``The Physics of Star Formation Feedback and Molecular Cloud Destruction" and an Alfred P. Sloan Research Fellowship.

\appendix
\section{Flux-Limited-Diffusion Algorithm}\label{section_FLD}

\begin{figure}
\begin{center}
\includegraphics[width=9.0cm]{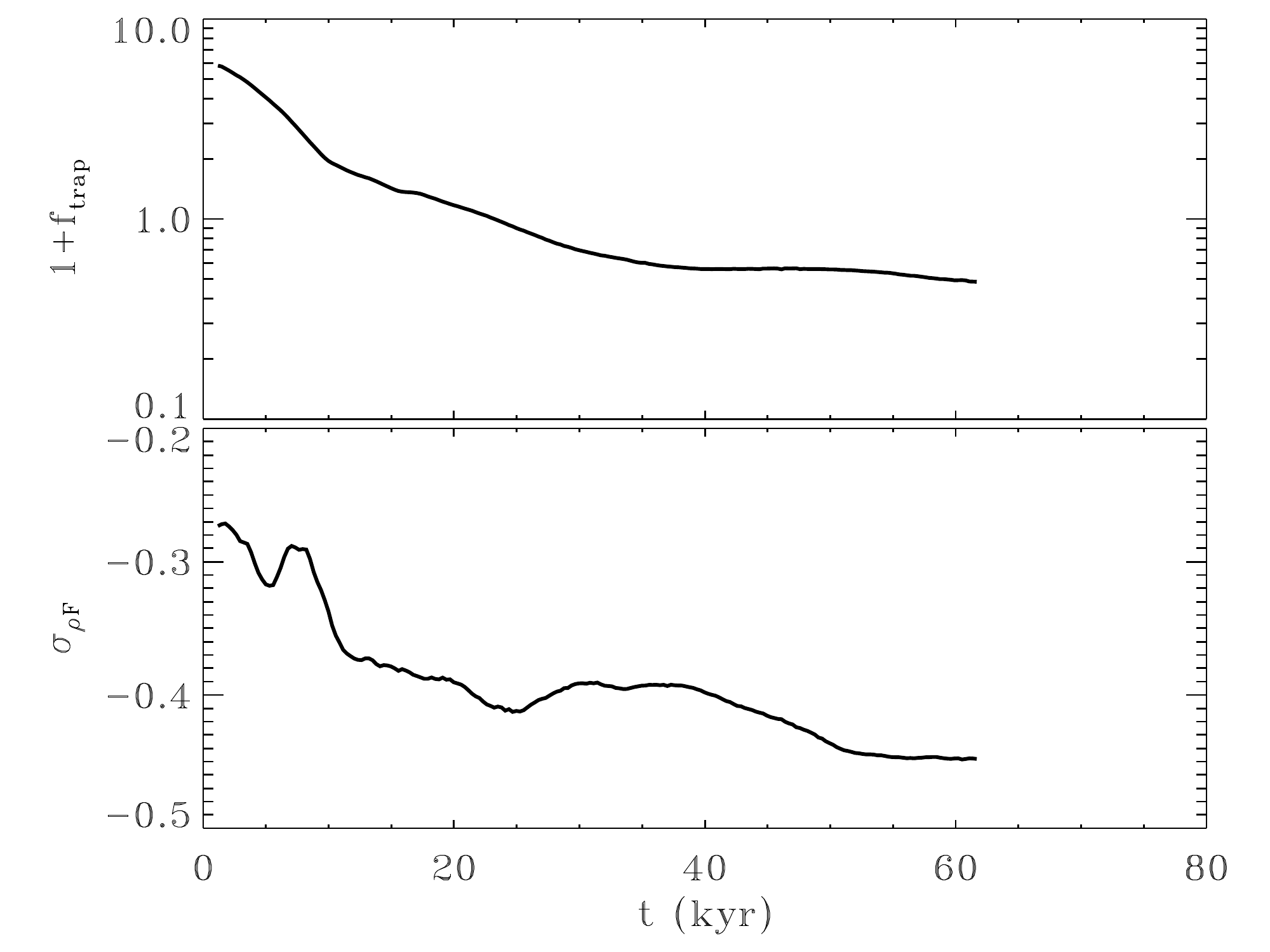}
\end{center}
\caption{Upper: trapping factor $1+f_{\rm trap}$ as a function of time for the FLD run. Lower: correlation between density and flux $\sigma_{\rho F}$ avarged over grid zones with $\rho \geq 10^{-6}\rho_{*}$ for the FLD run. }\label{fig_FLD}
\end{figure}

In order to isolate the effects of the radiation transfer algorithm, we carry out a single simulation with the \textsc{athena} FLD module to compare with our VET results.  We choose $\tau_* = 3$ with the computational box size $[L_x \times L_y]/h_a = 256 \times 2048$ and a high resolution $N_x \times N_y = 1024 \times 8192$. The boundary conditions match the FLD boundaries described in further detail in D14.  The horizontal boundary conditions are periodic.  The bottom vertical boundary is chosen to enforce a constant flux at the base and the top vertical boundary assumes a free-streaming limit. The initial condition for the FLD run is set up by the end state of the T3\_F0.5\_FLD run described in D14. The gas in T3\_F0.5\_FLD eventually reaches a quasi-steady state in a gravitational field, and we take the gas at $t=200\,t_*$ in T3\_F0.5\_FLD as the initial condition of our FLD run, then we turn off the gravity. Following D14 (see also KT12) we adopt a diffusion-like equation for the flux instead of the VET algorithm
\begin{equation}
\mathbf{F}_r=-\frac{c \lambda}{\sigma_{F}} \nabla E_r,\label{eq:fld}
\end{equation}
where the flux-limiter is given by 
\begin{eqnarray}
\lambda &= &\frac{1}{R}\left(\coth{R} - \frac{1}{R}\right)\nonumber \\
R &=&\frac{| \nabla E_r |+\beta}{\sigma_{F} E_r}.\label{eq:limiter}
\end{eqnarray}
The parameter $\beta$ is a small number as an effective floor on $R$ in an infrared optically thin region, it helps to obtain convergence  in the FLD code. We choose $\beta=4\times10^{-4}$, as given in D14.

We find that the gas is accelerated much more slowly than that in T3L and T3H runs in Section \ref{results}. We stop the run at $t= 25 t_a \sim 60$ kyr when the gas spreads out and almost fills out the entire computational box. Similar to KT13, we find that the FLD run forms many long and vertically oriented low-density channels where the flux can escape. The upper panel of Figure \ref{fig_FLD} shows the trapping factor in the FLD run. We still use the definition of $f_{\rm trap}$ in equation (\ref{trapping1}), but plot $1+f_{\rm trap}$ instead of $f_{\rm trap}$. Our results are qualitatively consistent with KT13, but we find the trapping factor drops even further than they do and the resulting momentum coupling is even weaker than they claim. Importantly, the clear differences between these results shown in Figure \ref{fig_FLD} and our VET results in Figures \ref{trapping} and \ref{correlation} provide a strong indication that the drop in trapping factor observed by KT13 is a radiation transfer dependent effect.

The lower panel in Figure \ref{fig_FLD} shows the volume weighted correlation between density $\rho$ and $y-$direction flux $F_{ry}$, excluding grid zones with density $\rho < 10^{-6}\,\rho_*$.  As far as the simulation ends, we find that $\sigma_{\rho F}$ also decreases with time. Some form of radiative RTI may be responsible for the trapping factor decreasing although it is unclear if the standard radiatve RTI would be present as a linear instability in this limit where $g \rightarrow 0$. As discussed in Section \ref{sec_trapping}, we conclude that raditive RTI, if present at all, has relatively little impact in VET simulations in a long term evolution.

%These results are different from the VET runs as shown Figures \ref{trapping} and \ref{correlation}, which show fairly flat values of $f_{\rm trap}$ and $\sigma_{\rho F}$. As discussed in \ref{sec_trapping}, we conclude that raditive RTI affects the long term evolution of momentum coupling between the gas and radiation as well as the flux-density correlation in FLD simulations, but has relatively little impact in VET simulations. 

\clearpage

\end{document}